\documentclass[fleqn,usenatbib]{mnras}

\usepackage{newtxtext,newtxmath}

\usepackage[T1]{fontenc}

\DeclareRobustCommand{\VAN}[3]{#2}
\let\VANthebibliography\thebibliography
\def\thebibliography{\DeclareRobustCommand{\VAN}[3]{##3}\VANthebibliography}

\usepackage{graphicx}	
\usepackage{amsmath}	

\usepackage{ulem}
\usepackage[dvipsnames]{color}

\def\be{\begin{equation}}
\def\ee{\end{equation}}
\def\ba{\begin{eqnarray}}
\def\ea{\end{eqnarray}}

\def\msun{M_\odot}

\def\ltsima{$\; \buildrel < \over \sim \;$}
\def\simlt{\lower.5ex\hbox{\ltsima}}
\def\gtsima{$\; \buildrel > \over \sim \;$}
\def\simgt{\lower.5ex\hbox{\gtsima}}

\definecolor{webgreen}{rgb}{0,.5,0}
\definecolor{webbrown}{rgb}{.6,0,0}
\definecolor{falured}{rgb}{0.5, 0.09, 0.09}

\title[Dust during the SN-ISM interaction]{Dust evolution in a supernova interacting with the ISM }

\author[Vasiliev \& Shchekinov]
{
Evgenii O. Vasiliev$^{1}$, Yuri A. Shchekinov$^{2}$  \\
\footnotesize \it $^{1}$Lebedev Physical Institute, Russian Academy of Sciences, 53 Leninsky Avenue, Moscow 119991;\\
\footnotesize \it $^{2}$Raman Research Institute, Sadashiva Nagar, Bangalore 560080, India;
}
\date{Accepted XXX. Received YYY; in original form ZZZ}

\pubyear{2015}

\begin{document}
\label{firstpage}
\pagerange{\pageref{firstpage}--\pageref{lastpage}}
\maketitle

\begin{abstract}
Supernovae (SN) explosions are thought to be an important source of dust in galaxies. At the same time strong shocks from SNe are known as an efficient mechanism of dust destruction via thermal and kinetic sputtering. A critically important question of how these two hypotheses of SNe activity control the dust budget in galaxies is still not quite clearly understood. In this paper we address this question within 3D multi-fluid hydrodynamical simulations, treating separately the SNe injected dust and the dust pre-existed in ambient interstellar gas. We focus primarily on how the injected and the { pre-existing dust is} destroyed by shock waves and hot gas in the SN bubble depending on the density of ambient gas. Within our model we estimate an upper limit of the SN-produced dust mass which can be supplied into interstellar medium. For a SN progenitor mass of 30 $\msun$ and the ejected dust mass $M_d=1~\msun$ we constrain the dust mass that can be delivered into the ISM as $\geq 0.13~\msun$, provided that the SN has injected large dust particles with $a\geq 0.1~\mu$m.
\end{abstract}



\section{Introduction} 

Dust is an important constituent of the interstellar medium (ISM), playing key role in physical processes that determine its basic properties: chemical transformations, metal budget in gas phase, thermodynamic state, radiation transfer. Dust is known to convert stellar light and thermal energy of hot gas in infrared (IR) radiation. One of the intriguing questions concerns the dust mass budget in the ISM \citep[see for recent discussion in ][]{Mattsson2021,Kirchschlager2022,Peroux2023}. The interrelation between the destruction and production dust rate in the ISM is still not quite well understood. { Of particular concern is the processing} the dust particles undergo behind the strong shock waves, $v_s\simgt 150$ km s$^{-1}$, penetrating the ISM. It is generally thought that dust particles experience efficient destruction from such shocks and from the hot gas behind them. The three processes dominate the dust destruction: the inertial and thermal sputtering \citep[][ and references therein]{Barlow1978,Draine1979a,Draine1979b,McKee1989,Jones1994,Nath2008,Slavin2015,Priestley2022}, and the shattering in grain-grain collisions at higher densities \citep[][]{Borkowski1995,Jones1996,Slavin2004,Guillet2009,Bocchio2016,Kirchschlager2019}. 

Theoretical considerations show that the characteristic dust lifetime in the Milky Way ISM against { sputtering is estimated to be}  $t_{sp}\simlt 3\times 10^8$ yr \citep[][]{McKee1989,Jones1994} to $t_{sp}\simlt 3\times 10^9$ yr \citep[][]{Jones1994b,Slavin2015}, resulting in the decrease rate $\dot M_d^-\simlt (0.1-0.01)\msun$ yr$^{-1}$; more recent discussion can be found in \citep[][]{Bocchio2014,Ginolfi2018,Micelotta2018,Ferrara2021}. On the other hand, the overall production rate from red giants winds and SNe explosions is $\dot M_d^+\sim 10^{-3}\msun$ yr$^{-1}$, where ${\rm  SFR}\sim 5\msun$ yr$^{-1}$ is assumed, indicating a { severe}  disbalance between the dust destruction and its replenishment \citep[][]{Draine2009,Bocchio2016}. This discrepancy can be mitigated when dust-to-gas decoupling under the action of shock waves and gravity for large particles is accounted for \citep[][]{Hopkins2016,Mattsson2019,Mattsson2022}. However, one to two orders of magnitude difference between the $\dot M_d^-$ and the $\dot M_d^+$ requires apparently a proportional amount of dust to be hidden of destructive SNe shocks, which seems unrealistic. An additional and apparently efficient dust mass supply can be connected with growth of grains in the ISM \citep[][]{Draine1990,Chokshi1993,Dwek1998}, and more recently \citep[][  and references therin]{Calura2008,Draine2009,Mattsson2011,Inoue2011,Ginolfi2018,Heck2020}. Moreover, supersonic turbulence is shown to be an efficient mechanism that can stimulate formation of dust in the ISM of local galaxies and even in the early Universe, and as such can counteract  { efficient} dust destruction by SNe \citep{Hopkins2016,Mattsson2019,Mattsson2019b,Mattsson2020,Mattsson2020b,Li2020,Commercon2023}.

The discrepancy between the production and destruction rate { is seen in particular} in galaxies at high redshifts as first pointed out by \citet[][]{Todini2001,Morgan2003}. The detection of dust in quasars at $z\sim 5$ (the universe's age $<1$ Gyr) suggests  the SNe II to be the { dominant} dust source in the early universe \citep[][]{Bertoldi2003,Maiolino2004,Beelen2006,Valiante2009,Valiante2011,Dwek2011,Gall2011,Riechers2013}. Moreover, further observations { at} intermediate redshifts $z\sim 1-5$ with the {\it Herschel Space Observatory} revealed a more generic problem -- an apparent excess of dust in submillimeter and ultraluminous IR galaxies -- the so-called {\it `dust budget crisis'} \citep[][]{Michalowski2010,Dunne2011,Rowlands2014}. 

Direct IR observations in the local Universe also show that dust can be produced at early stages of the ejecta outflows as manifested in several nearby SN remnants \citep[in particular, in Cas A and Crab, ][]{Dunne2003,Gomez2012,Arendt2014}, in SN1987A \citep{Indebetouw2014,Matsuura2015}, and in the local group galaxy NGC 628 \citep{Sugerman2006}. More recent analysis of IR characteristics from SN1987A \citep{Wesson2021} indicates that dust forms in SNe ejecta at latter stages $\simgt 1000$ days. The latter may reflect the fact that the net dust product by SNe is environmentaly sensitive as demonstrated by \citet[][]{Nozawa2006}.

Recent measurements of abundance patterns in supernovae remnants (SNR) in the Milky Way (MW) and Large Magellanic Clouds (LMC) with strong non-radiative shocks, $v_s\simeq 350-2700$ km s$^{-1}$, have also constrained the destroyed dust fraction by $\simlt 0.1-0.6$, even in a rather dense environment \citep[see Tables 1 and 3 in ][]{Zhu2019}. { It is therefore conceivable} that older SNRs with weaker shock waves are less destructive than commonly thought. More recent observations of three SNRs in the MW { do confirm} such a conclusion \citep[][]{Priestley2021}.
 
The total dust mass supplied by a SN in the ISM, and the dust size distribution sensitively depend not only on the progenitor mass, but on the density of ambient gas as well, because of the reverse shock from the interaction of the ejecta with the ambient gas. 1D simulations by \citet[][]{Nozawa2006,Nozawa2007,Bianchi2007,Nath2008} and \citet[][ for more recent discussion]{Bocchio2016} have shown that increase of the ambient gas density from 0.1 to 10 cm$^{-3}$ can result in a drop of dust yield by one to two orders, particularly for higher progenitor masses. { Equally} important is the conclusion that destruction of small size dust by the reverse shock considerably flattens the size spectrum. More recently, theoretical analysis of dust destruction in the process of interaction of the ejecta with ambient ISM leads \citet{Slavin2020} to conclude that { when relative motion} of dust particles with respect to { the} gas is accounted, the heavier dust particles can penetrate the region affected by the reverse shock and can escape relatively intact into the ambient ISM gas. However, eventually a fraction of them is becoming destroyed {\it in situ} in the surrounding ambient gas. This result suggests that { SNe inject mainly large dust particles into the ISM}. This { in turn} can contribute to variations of the extinction law on smaller scales, unless dust particles are well mixed with dust in ambient gas. 

In this paper we present { results} of 3D multi-fluid hydrodynamical simulations of how { SN} ejecta with dust particles having formed in the initial stages of its evolution propagates { through the dusty interstellar} gas. We { concentrate on} the question of how dust particles injected by the SN and those present in the surrounding ISM are destroyed by strong shock waves. The injected dust particles and the interstellar dust are treated as two different particle populations experiencing distinct evolutionary paths. Formation and growth of dust particles in dense cool regions of the shell surrounding the remnant are not included in our consideration. Section \ref{sec:model} sets up our model. Section \ref{sec:dynamics} describes the obtained results, including dynamical aspects related to dust destruction and overall mass budget of the injected dust and the dust belonging to ambient ISM (refereed as the { interstellar or pre-existing} dust). Section \ref{sec:discus} contains a general discussion and the summary.

\section{Model description} \label{sec:model}

We consider the dynamics and destruction of dust particles in a supernova remnant during its {300 kyr-long} interaction with a slightly inhomogeneous {clumpy} ISM. We use our gasdymanic code \citep{vns2015,vsn2017} based on the unsplit total variation diminishing (TVD) approach that provides high-resolution capturing of shocks and prevents unphysical oscillations, and the Monotonic Upstream-Centered Scheme for Conservation Laws (MUSCL)-Hancock scheme with the Haarten-Lax-van Leer-Contact (HLLC) method \citep[see e.g.][]{Toro2009} as approximate Riemann solver. This code has successfully passed the whole set of tests proposed in \cite{Klingenberg2007}. {In order to} follow the dynamics of dust particles we have implemented the method similar to that proposed by \citet{Youdin2007}, \citet{Mignone2019} and \citet{Moseley2023}. A description and tests are given in Appendix. In this paper we { only take} destruction of dust particles by both thermal (in a hot gas) and kinetic (due to a relative motion between gas and grains) sputtering \citep{Draine1979b} { into account}. In order to separate different mechanisms that can contribute to dust processing, we do not consider here additional effects from possible growth of dust particles in denser regions of the remnant, and their fragmentation from shattering collisions. These processes operate on time scales of a few Myr \citep[see, e.g., ][]{Hirashita2011,Mattsson2020}, even in much denser environments { (molecular clouds)} $n\simgt 10^2$ cm$^{-3}$ \citep[][]{Martinez2022}. { This is} much longer than the ages of SNe remnants in a diffuse ISM with the shocks  sufficiently strong ($v_s\simgt 150$ km s$^{-1}$) for dust destruction: $t\simlt 100$ kyr.

The backward reaction of dust on to gas due to momentum transfer from dust particles is also accounted in order to ensure dynamical self-consistency. Generally dynamical effects from dust are minor, however in the post-shock domain, such as the postshock shell, the collisional coupling is weak, and dust particles move inertially with velocities $v_d\geq 3v_s/4$, $v_s$ is the shock velocity. As a result, their contribution to ram pressure $\rho_dv_d^2$ cannot be negligible until the dust velocity relaxes to the gas one $|{\bf v}_g-{\bf v}_d|\ll c_s$, $c_s$ being the sound speed (for further discussion see Sec. \ref{intermx}). In addition, the effects of backward reaction can be of importance in correct estimates of the kinetic sputtering for particles inertially entering the ambient gas at the interface between the bubble interior and the surrounding gas.

The effects of magnetic field are apparently of critical importance for dynamics of dust particles in SN remnants, particularly in terms of their destruction. It is known, that in presence of { betatron acceleration} of charged dust can enhance the efficacy of sputtering at  radiative stages on an expanding SN remnant, however, only when dust particles are decoupled of gas \citep[][]{Shull1977,Draine1979b,Seab1983,Slavin2004}. As { we show below} (Sec. \ref{intermx}), even massive particles, $a\geq 0.1~\mu$m, lose collisional coupling only behind the shock front in outer regions of the hot bubble where radiation cooling is insignificant, whereas smaller ones remain collisionally linked to gas. In addition, as shown by \citet[][]{Slavin2004}, individual trajectories of dust grains are very sensitive to grains' sizes and charges, their chemical composition and shock velocity with gyration radii of 0.03 to 0.3 pc. Adequate numerical description on scales of SNe remnants requires an unprecedented resolution, and is worth separate consideration.      

The gas distribution in the ambient (background) medium is set to {be} a slightly inhomogeneous with averaged number density $n_b$ and {small-size low-amplitude} density fluctuations {$\delta r\sim 2-3$ pc,} $\delta n /n_b \sim 0.1$. The fluctuations are constructed by using the module pyFC\footnote{The code is available at https://bitbucket.org/pandante/pyfc/src/master/} by \citet{Lewis2002} which generates lognormal ``fractal cubes'' for gas density field. The averaged ambient density is equal to $n_b = 1$~cm$^{-3}$ as a fiducial value, we calculate also models with 0.1, 0.3, 3 and 10~cm$^{-3}$, the temperature in all models is $10^4$K. Initially the gas pressure is assumed uniform { over the whole} computational domain. The metallicity of the background gas is usually set to be solar ${\rm [Z/H]} = 0$. We assume a dust-to-gas (DtG) mass ratio equal to $\zeta_b=10^{-2-{\rm [Z/H]}}$. 

We inject the mass and energy of a SN in a region of radius $r_0=3$~pc, assuming commonly used values { of mass and energy}$30~\msun$ and $10^{51}$~erg. The energy is injected in thermal form. The injected masses of metals and dust are $10 ~\msun$ and $1~\msun$, correspondingly. Dust particles are thought to start growing during the free expansion phase of a SN evolution on much earlier times, when the ejecta temperature drops to several thousand Kelvins \citep[e.g.][]{Dwek1980,Todini2001}, typically after $\simgt 300$ days when the ejecta radius is of $\sim 0.03$ pc. At this age the characteristic time of dust-gas collisional coupling (the stopping time, see Sec. \ref{dyn_def}) for the parameters within the ejecta is $\tau_s\sim 10^6n^{-1}a_{0.1}$ s, with $a_{0.1}$ being the dust grain radius in $0.1~\mu$m. For $M_{0}=30~\msun$, and $r_0=3$ pc, the ejecta gas density $n\sim 10^3$ cm$^{-3}$, resulting in $\tau_s\sim 10^3a_{0.1}$ s at gas temperatures typical for dust growth. The dust grains nucleate and grow {\it in situ} in the ejecta as first described by \citet{Dwek1980}, and have developed further on in \citet{Todini2001}. This would suggest that the formed dust particles { get mixed into the ejecta} on a short time scale. At such conditions in the process of acceleration the relative velocity of dust grains and gas is small, and only a small fraction of dust particles experiences kinetic sputtering \citep[][ see also estimates in Sec. \ref{app_dest}]{Slavin2020}. The thermal sputtering remains apparently weak because of a low temperature in ejecta $T<6000$ K. { However, a firm estimate of the surviving dust fraction requires more detailed consideration}. In this sense the estimates of the contribution of dust supply into the ISM from SNe explosions presented in our model can be regarded as upper limits. When the ejecta meets the reverse shock and becomes thermalized at $t\sim 300$ yr, thermal sputtering increases and comes into play  \citep{Nozawa2007,Nath2008,Slavin2020}. Overall, at $t\geq 300-400$ yr the destroyed dust mass fraction can vary from 0.3 to 0.8 depending on { dynamics} in the surrounding gas \citep{Slavin2020}.   

The mass of dust is redistributed between many dust 'superparticles'. For instance, in case the number of dust 'superparticle' is  $2^{20}\sim 10^6$, the mass of each dust 'superparticle' is $\sim 1\msun/10^6 \sim 10^{-6}\msun$ for the monodisperse dust of total mass equal to $M_i=1\msun$. In the equation of motion of dust we attribute to a single grain the mass and the size of physical dust particles, while in the momentum equation for the gas component the momentum from dust is treated as a sum of momentum from all dust grains that constitute the dust mass $\sim 10^{-6}~\msun$ allocated in a cell.

In this paper we assume dust grains to have equal sizes ({\it monodisperse} dust model) { at} $t=0$. We perform a set of independent runs for { monodisperse dust populations} with initial sizes equal to $a_0 = 0.03, \ 0.05, \ 0.1, \ 0.3, 0.5\mu$m; $a_0=0.1\mu$m is the fiducial value. During the evolution dust grains are destroyed and their sizes decrease depending on physical conditions in ambient gas; the minimum size is set to $0.01\mu$m. We adopt the same initial distribution of sizes both for the dust pre-existed in ambient gas (referred hereinafter as the { interstellar} dust) and for the injected dust as well. We assume that the mass of metals produced due to sputtering is returned to the gaseous phase. Evolution of polydisperse dust will be described elsewhere. 

In our models the spatial resolution is set to 0.5~pc, that is sufficient for adequate treatment of dynamics of a SN bubble in a medium with  density $n_b \sim 0.1-10$~cm$^{-3}$. In particular, in a medium with $n_b = 1$~cm$^{-3}$ we follow the SN bubble evolution until late radiative phase: $t_{end} \sim 10 t_{cool} \sim 300$~kyr, when its radius reaches around $\simeq 40$pc. Therefore, we set the size of the computational domain ($96$~pc)$^3$, with the number of cells $192^3$. We set one dust 'superparticle' per a computational cell resulting in the total number of { interstellar} dust particles in the domain $192^3 \sim 7$ millions. The fiducial number of injected 'superparticles' per SN is $2^{23} \sim 8$ millions. Our choice of both grid resolution and number of the injected particles is justified by our tests presented in Appendix~\ref{sec:gentest}. { Note that} dust particles can escape their initial ``mother'' cells and move into { neighbouring ones}. As a result, depending on gas density the overall distribution of dust particles can be rather patchy and can { lead to local deficiency of dust particles, also seen in} averaged radial profiles, see in Sec. \ref{decoupl}. 

Simulations are run with tabulated non-equilibrium cooling rates fitting the calculated ones for a gas that cools isochorically from $10^8$ down to 10~K \citep{v11,v13}. The heating rate is assumed to be constant, with a value chosen such as to stabilize the radiative cooling of the ambient gas at $T=10^4$~K.

\section{Dynamics} \label{sec:dynamics} 

\subsection{Dust radial distribution}\label{radpro}

Inside a SN bubble dust is efficiently destroyed by both the thermal and kinetic sputtering \citep{Draine1979b}. { The effect of the latter on interstellar dust} particles in a { unmagnetised} case is around 15\%, while for the injected ones it is even smaller because of a lower drift between the gas and dust\footnote{Details are given in Appendix \ref{app_dest}.}. Fig.~\ref{fig-2d-slices} presents angle averaged radial profiles for gas density and temperature, and densities of both sorts of dust particles -- the SN  injected and the { interstellar}  dust, for the fiducial grain size at 50 and 300 kyr after the SN has exploded. Because of initially introduced clumpy density distribution and partly because of { growing Rayleigh-Taylor (RT) instabilities} the averaged shock layers in $n(r)$ and $T(r)$ are wider than typical. At $t=50$ kyr the injected dust remains $\sim 7$ pc deeper behind the shock and { about 10\% survive} within the central $\sim 10$ pc interior. The { interstellar} dust is swept by the shock and penetrates behind it by $\sim 5$ pc inwards, but only in a $\sim 2-3$ pc thick layer it survives within factor of 2, as seen from comparison of the { magenta} (with sputtering accounted) and the { light-magenta} (no sputtering is included) lines in upper paned of Fig.~\ref{fig-2d-slices}. At 300 kyr only a small fraction of injected dust survives within the bubble interior, around 10\% if it penetrates ahead of the forward shock { as shown by the line representing interstellar dust which has already crossed the shock}, $n_{d,e,{\rm SN}}$, and partly experienced sputtering. They apparently experience  acceleration episodes from the RT ``tongues'' that advance the shock front and penetrate outward\footnote{This is also observed in the model by \citet{Slavin2020}.}. Behind the front their density drops within { a thin layer of $\delta r\simlt 5$ pc},  as can be observed in lower panel of Fig. \ref{fig-2d-slices}.

\begin{figure}
\includegraphics[width=8cm]{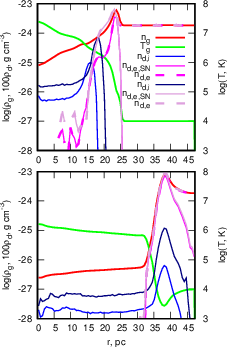}
\caption{
The angle averaged radial profiles of gas density ($n_{g}$, red line), gas temperature ($T_{g}$, green line), density of the injected dust { (``i'')} particles ($n_{d,i}$) with sputtering (blue line) and without it (dark-blue line), density of the { interstellar (pre-existed, ``e'')} dust particles ($n_{d,e}$) with sputtering ({ magenta} dash line) and without it ({ light-magenta} dash line) and density of the { interstellar} dust particles behind the SN shock front ($n_{d,e,SN}$) with sputtering ({ magenta} solid line) and without it ({ light-magenta} solid line) at 50 and 300~kyrs (upper and lower panels, respectively). The initial grain size of the injected and { interstellar} particles is equal to the fiducial value: $a_0 = 0.1\mu$m.
}
\label{fig-2d-slices}
\end{figure}

\subsection{Dust-gas decoupling}\label{decoupl}
\subsubsection{Spatial dust distribution pattern around the shell}

\begin{figure*}
\includegraphics[width=18.0cm]{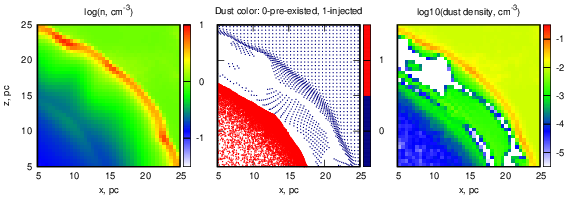}
\caption{
A part of the SN shell at the radiative stage: 2D slices of the gas density (left), positions of dust particles (middle), average dust density (right) at 50~kyrs. The initial grain size is equal to the fiducial value: $a_0 = 0.1\mu$m.
}
\label{fig-den-2dslice}
\end{figure*}

Dust particles are initially { placed on the} computational grid as described above in Sec. \ref{sec:model} and in Sec. \ref{prop}. { Dust particles can escape their original ``mother'' cells} and move into neighbour ones, in those cases when the stopping time $\tau_s$ becomes longer than the crossing time for a grid cell $t_{\rm sound}$, and the collisional coupling weakens, { see} Sec. \ref{intermx}. As a result, depending on gas density and number of particles the overall distribution of dust particles can become  patchy, { with a lack or deficiency of} dust particles at certain regions of the bubble and in the averaged radial profiles. Fig.~\ref{fig-den-2dslice} { shows the} distribution of gas and dust density in and around the remnant at the radiative stage. The middle panel presents the positions of dust particles in a 2D slice (a single cell in thickness) at a certain time after explosion. Initially the { interstellar} dust particles are deposited to the computational grid homogeneously with a density of one particle per a cell in its centre, as seen in Fig. \ref{fig-den-2dslice} in the unperturbed region. Evacuation of dust particles due to decoupling are clearly seen in several regions behind the front. Within the hydrodynamical description dust particles along with their physical characteristics mentioned in Sec.~\ref{sec-deposit} are smeared out over the computational cell resulting in mitigation of this patchiness. { This results in reducing the area} with lacking dust seen in the right panel of Fig.~\ref{fig-den-2dslice}, as compared with the { pattern} shown in the middle panel. The radial profiles shown in Fig.~\ref{fig-2d-slices} { include} this effect, see also discussion in Sec.~\ref{intermx}. In this regard it is also worth noting that a relatively small number of ``superparticles'' -- one per computational cell, may cause numerical artefacts in the immediate post-chock domain, where { dust and gas decouple} (see Fig.~\ref{fig-size-rad} in Sec.~\ref{intermx}). The smearing out procedure partly mitigates possible contaminations from such artefacts.

{ In the left panel we can see} a tendency of gas density to vary quasi-periodically along the thin shell. Similar signs of quasi-periodicity can be found also in { the} distribution of dust ``superparticles'' in the middle panel. These signs { indicate} development of both the thermal instability and the Vishniac{--Ostriker} overstability \citep[][]{Ostriker1981,Vishniac1983,White1990} at their initial stages. It is difficult though to { tell them apart at early times}, whereas at later stages the thermal instability results in fragmentation of the shell and seems to hinder further development { of} the Vishniac{--Ostriker} overstability. However, the shell expansion at $t\simgt 50$ kyr follows the standard Oort law $R_s\propto t^{1/4}$, in which case the growth rate of the Vishniac{--Ostriker   overstability} $s\simeq 0$ \citep[see Eq. 2.17d in ][]{Vishniac1983}. More recent on-puprose numerical study of the { Vishniac--Ostriker overstability} by \citet{Miniere2018} shows that perturbations from { the interior regions cause the instability} to reappear at the radiative phase. In our case during the period from 50 to 300 kyr we do not observe signatures of the Vishniac{--Ostriker overstability} in the cooled shell while it becomes denser. The discrepancy can be due to differences in the initial perturbations introduced into the computational domain. However, it is worth noting that effects from possible growth of the Vishniac{--Ostriker overstability}  cannot affect dust destruction.

\subsubsection{Pre-existing dust} 

\begin{figure}
\includegraphics[width=8.5cm]{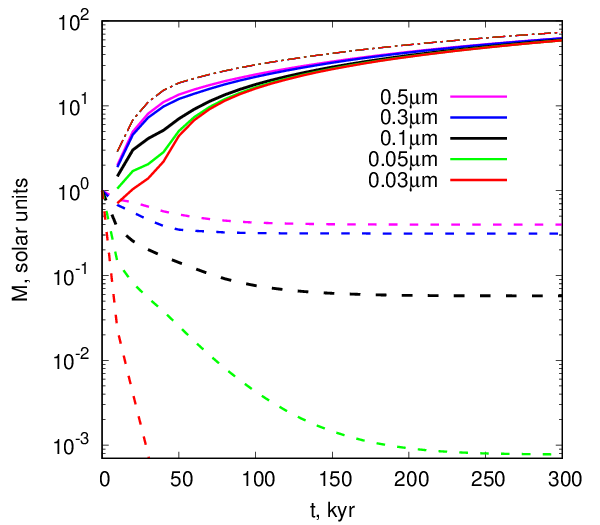}
\caption{
The total mass of dust involved into the SN bubble at a given time. Solid lines show the evolution of the { interstellar} dust (the dust pre-existed in the ambient gas), dot-dashed red line shows the mass of the same dust without destruction, with the asymptote $M_d\propto R_s^3\propto t^{3/4}$. Dashed lines depict the evolution of the injected dust; the injected dust mass is $M_d=1~\msun$ independent on grain initial radius $a_0$.
}
\label{fig-evol-mass}
\end{figure}

The { interstellar} grains in the swept up gas of the SN shell are destroyed efficiently during the early expansion $t\simlt 50$ kyr. This can be observed in { the} upper panel ($t=50$ kyr) of Fig. \ref{fig-2d-slices}, where the dust destruction { is lower} in the immediate post-shock region, whereas { in the hotter innermost} domain it remains considerable. The reason is that dust particles of radius $a\simgt 0.1~\mu$m are destroyed { on a timescale of} $t\sim 10^2n^{-1}a_{0.1}$ kyr comparable to radiative cooling time $t_{\rm cool}\sim 10^2n^{-1}$ kyr. The remnant enters the radiation dominated phase at $t\sim 30$ kyr, and as a consequence dust destruction rate { decreases} with decreasing temperature. Solid lines in Fig.~\ref{fig-evol-mass} depict the mass of the { interstellar} dust { included in} the SN bubble. On longer { timescale,} sputtering rate falls below the necessary level, and asymptotically the swept up dust mass grows proportionally to $\propto R_s^3\sim t^{3/4}$. For comparison, { the} dot-dashed red line shows the dust mass density profiles for the case without dust destruction. 

\subsubsection{Injected dust}

Dashed lines in Figure~\ref{fig-evol-mass} show evolution of the total mass of { dust populations of various initial sizes} $a_0$ at the time of injection $t=0$ by the SN. Small injected particles are destroyed quickly with characteristic sputtering time $\propto a$, such that grains of initial size $0.05\leq\mu$m are mostly sputtered { away} within 10~kyr after SN explosion. During { the subsequent} evolution the shock velocity and the gas temperature in the bubble and in the ejecta decrease, and consequently the efficiency of dust destruction also decreases, as can be judged from dust radial profiles on lower panel of Fig. \ref{fig-2d-slices}. The apparent increase of the { interstellar} dust mass at later stages, $t>100$ kyr, is due to the swept up dust at the Oort expansion phase, $M_d\propto R_s^3\propto t^{3/4}$. { As a result, sputtering of injected dust particles of all sizes levels out at} $t\sim 100$ kyr. { In the end,} the mass of { surviving} dust is $\sim 5$\% for particles of $a_0=0.1~\mu$m to $\sim 40$\% for $a_0=0.5~\mu$m. 
 
\begin{figure}
\includegraphics[width=8.5cm]{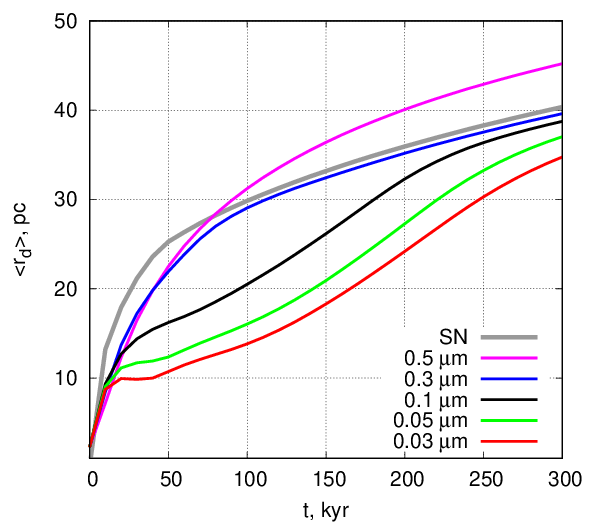}
\caption{
The averaged distance $\langle r_{a_0}(t)\rangle$ of ensembles of the dust particles from the point of their injection, i.e. the point of a SN explosion. The value of $\langle r_{a_0}(t)\rangle$ is weighted to the current particle's supermass $m_i$: $\langle r_{a_0}(t)\rangle = \sum{r_i m_i}/\sum{m_i}$.  The particles have been injected at $t=0$ with the initial radii from $a_0=0.03~\mu$m to $0.5~\mu$m as shown in the legend. Color lines mark $\langle r_{a_0}(t)\rangle$ for an ensemble corresponding to a given $a_0$; the thick grey line depicts the radius of the SN bubble.
}
\label{fig-evol-rd}
\end{figure}

Before { going into details about} dust destruction on smaller scales, let us consider first more generic averaged behavioural features. This gives us an impression { of} a collective dynamics of the injected particles depending on their initial radii $a_0$. For this we calculate the averaged { distance $\langle r_{a_0}(t)\rangle$ of dust particles injected by the SN at $t=0$ with radius $a_0$}. At $t\simgt 100$ kyr the SN remnant { reaches} the radiative stage and forms a relatively dense and thin shell. The injected dust particles, { which are initially surrounded by} the hot gas of the ejecta are destroyed mostly by thermal sputtering. The fraction of the remnant volume, { that is} occupied asymptotically at $t\geq 100$ kyr, depends on the particles' initial size. Fig. \ref{fig-evol-rd} illustrates the evolution of $\langle r_{a_0}(t)\rangle$ for different $a_0$. At early times ($\simlt 10$~kyr), during the free expansion and and Sedov-Taylor phases, $\langle r_{a_0}(t)\rangle$ { shows a weak dependence} on $a_0$. However, it changes further on: small particles, $a_0<0.1~\mu$m, are tightly coupled to the ejecta and the hot bubble interior, and remain within the inner 30--36 pc of the bubble. During the next $\sim 200$~kyr such particles along with the hot ejecta overtake the dense shell and {remain inside the it}. Heavier dust particles, $a_0\geq 0.1~\mu$m, reach either the post-shock layer, or even overcome it, as $a_0=0.5~\mu$m particles do.

\subsection{Processing of dust}
\subsubsection{Dust survival vs gas ambient density} 

\begin{figure}
\includegraphics[width=8.5cm]{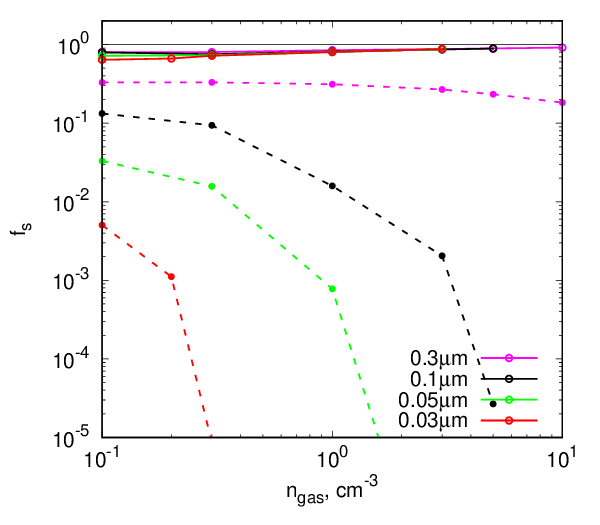}
\caption{
Dependence of survival fraction of the { pre-existing} (solid lines) and injected (dashed lines) dust at the asymptotic $t\geq 100$ kyr versus the ambient gas density; the horizontal thin solid line is the fraction without destruction.
}
\label{fig-survive}
\end{figure}

{ The surviving} fraction of the injected dust with a growing ambient density is shown by dashed lines in Fig. \ref{fig-survive}. As the ambient density increases the expansion velocity of the forward SN shock { decreases}, and as a consequence so does the expansion velocity of the ejecta, { thus prolonging the exposure of the injected dust to the hostile environment}. Even though the SN shell enters the radiation dominated phase in a  shorter time, it is sufficient to boost destruction of dust within the ejecta.

{ The interstellar} dust is much less sensitive to the ambient density as shown in Fig. \ref{fig-survive} by solid lines. This is connected with the fact that at gas temperature $T>10^6$ K for $a\sim 0.1~\mu$m the sputtering time $\tau_{\rm sp}\sim 10^5n^{-1}a_{0.1}$~s, is nearly equal to the gas cooling time, $t_{\rm cool}\sim 10^5n^{-1}$s at $T\sim 3\times 10^6$K, with $Z\sim Z_\odot$. For dust particles of $a\geq 0.03~\mu$m the sputtering time is only $\leq 3$ times shorter than cooling time. Below $T<10^6$ K the sputtering rate { decreases} as $\sim T^3$, resulting in simultaneous { reduction} of dust destruction. As the gas cooling rate { decreases} later, at $T\simlt 10^5$ K, { it is possible that the sputtering rate levels out before} gas cools considerably. As the SN remnant becomes radiative at $t\geq 50n^{-1}$ kyr, the independence of the survival yield of the { pre-existing} dust shown in Fig. \ref{fig-survive} is consistent with this scenario. Moreover, an apparent increase of the { surviving} fraction of the { interstellar} dust can even be observed at ambient density $n>0.2$ cm$^{-3}$. { This is related to the higher efficiency of} radiative cooling at higher densities and faster transition of the shock velocity to the Oort law $v_s\propto t^{-3/4}$. Similar { results} has been recently described by \citet{Kirchschlager2022}. In this regard it is worth noting that the survival yield shown here is already established within the intermediate asymptotic at $t\simgt 50n^{-1}$ kyr.
  
\subsubsection{Time variation of dust sizes}\label{timevar}

\begin{figure}
\includegraphics[width=7.1cm]{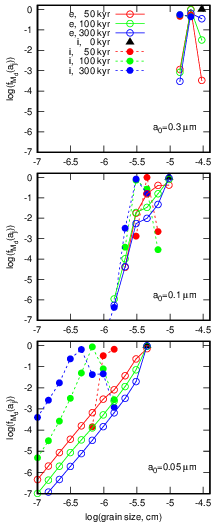}
\caption{
The mass distribution functions of grain sizes for the injected (filled circles connected by dashed lines, { ``i''}) and the { interstellar} (open circles connected by solid lines, { ``e''}) monosized particles of initial (at $t=0$) sizes: $a_0 = 0.3$, $0.1$ and $0.05\mu$m from top to bottom panels, respectively, in the SN bubble with age of 50, 100 and 300~kyr (red, green and blue lines). For $a_0 = 0.3\mu$m (top panel) the distributions of the injected particles at $t=100$ and $t=300$ kyr coincide. 
For the injected dust the initial distribution is marked by the black solid triangle symbol. The average number density of the ambient gas is 1~cm$^{-3}$.
}
\label{fig-pdf-size}
\end{figure}

Figure~\ref{fig-pdf-size} { shows the evolution} of the grain size { distributions} by mass inside the SN bubble for the injected and { interstellar} particles of different initial sizes: $a_0=0.3$, $0.1$ and $0.05\mu$m. The distribution function is defined as the mass fraction of grains in a given size bin $\Delta=a_j/a_{j-1}$ as
\be 
f_{M_d}(a_j) = {M_d(a_j)_{r<R_s} \over \Sigma_i M_d(a_j)_{r<R_s} },  
\ee
{ where} $R_s$ is the bubble radius, i.e., the radius of the forward shock, log$\Delta=0.165$ { with the number of bins being} $N=15$ within $0.01-0.3\mu$m.
Large grains are destroyed slower than the small ones proportionally to their radius $t_{\rm sp}\propto a$. 
Consequently, within the first 50 kyr the injected particles with $a_0=0.3~\mu$m decrease in size by less than { a factor of 2} and stay unchanged { beyond} $t = 100$~kyr. This is consistent with the total mass of { surviving} particles shown in Fig.~\ref{fig-evol-mass}. 

The { interstellar}  particles that crossed the front are destroyed less { efficiently}, because they spend in the postshock hostile region a shorter time, and only a negligible fraction pervade into the hot bubble and the ejecta. This is connected with the fact that at $t>50$ kyr the postshock gas becomes radiative, gas temperature falls below $T\simlt 10^6$ K and its density increases correspondingly. As mentioned above, the sputtering rate $|\dot a|/a\propto T^{3}$ decreases faster than the cooling rate $|\dot T|/T\propto T^{-1}$, { which leads to a decrease in the dust destruction rate in the shell}. The dust particles in their turn remains collisionally coupled to the gas, $\tau_s/t_{sound}\simeq 1$, and after passing a few cells behind the front they { remain in} the denser and cooler shell gas. After $\sim 50$ kyr only a minor fraction of the { interstellar} dust experinces sputtering. This explaines a { slow} growth of { the surviving} fraction of { the interstellar dust}.

The injected dust grains of smaller initial sizes $0.1$ and $0.05\mu$m are destroyed on { a} shorter time scale $\tau_{\rm sp}\propto a$. The destruction is { more efficient} during the very initial phase $t\simlt 50$ kyr when the ejecta is denser and hotter. Within $\sim 50-100$ kyr the maximum of $f_{M_d}(a_i)$ shifts by factor 2 toward smaller radii. On longer { timescales} between 100 and 300 kyr the sputtering rate { decreases} slightly because of a lower temperature in the ejecta. 

\begin{figure}
\includegraphics[width=7.5cm]{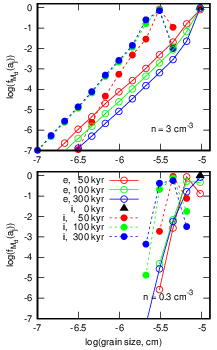}
\caption{
The same as in Figure~\ref{fig-pdf-size}, but for particles of initial size $a_0 = 0.1\mu$m and mumber densities of the ambient gas 3~cm$^{-3}$ (upper panel) and 0.3~cm$^{-3}$ (lower panel).
}
\label{fig-pdf-n}
\end{figure}

Increase of the ambient gas density enhances dust sputtering. { The resultant net effect} can be qualitatively described as a shift of the size distribution function $f_{M_d}(a)\sim f_{M_d}(a/n)$ at a given time, as can be found from comparison of the distributions in Fig.~\ref{fig-pdf-n} with the corresponding curves on Fig.~\ref{fig-pdf-size}.

\subsection{Purity of the { interstellar} and injected dust}\label{intermx}

The overall dynamics of dust destruction in either cases -- the { interstellar} and the injected particles, is predominantly determined by the interrelation between the collisional coupling (stopping) time $\tau_s$, the sputtering time $\tau_{\rm sp}$ and the local { sound-crossing} time $t_{\rm sound}=\Delta x/c_s$.  
{ The stopping time is roughly} $\tau_s\sim \rho_ma/\rho_g\sigma_t$ (Sec. \ref{dyn_def}), the thermal sputtering time at $T\simgt 10^6$ K is $\tau_{\rm sp}\sim\rho_ma/n_gm_T\sigma_t$, $m_T$ is the mass of a target particle \citep{Draine1995}. As can be inferred from Fig. \ref{fig-2d-slices} the ratio $\tau_{\rm sp}/\tau_s\sim m_{\rm H}/m_T\simlt 0.1$ for $m_T\simgt 10$. 

\begin{figure}
\includegraphics[width=8.5cm]{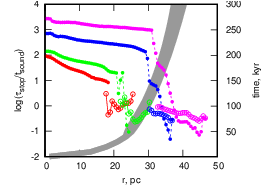}
\caption{
The radial profiles averaged over the solid angle for the size avegared ratio of the grain stopping time of the injected and { interstellar} dust to the local sound time $t_{\rm sound}$, $a_0=0.1\mu$m. Solid lines with open symbols are for the { interstellar} dust, dashed lines with filled symbols are for the injected particles. Shaded band represents evolution of the bubble shell: the right border corresponds to the forward front where the gas velocity jumps, and the left one shows the border where the gas temperature $T\geq 10^5$K. At $t\simlt 30-40$ kyr the SN remnant 
passes the initial period of the radiative phase, the borders practically coincide; the time is shown on the right $y$-axis. The { interstellar} particles lose collisional coupling starting from entering the shell and several (up to 10 pc) deeper inward. Color coding for the profiles corresponds to different ages: 50 kyr (red), 100 kyr (green), 200 kyr (blue) and 300 kyr ({ magenta}). 
}
\label{fig-size-rad}
\end{figure}

At early times $t\leq 50$ kyr the { interstellar pre-existing} particles { decouple from the gas shell and escape, since} $\tau_s/t_{sound}\simgt 1$. { this ratio decreases as} dust passes through the shell $\tau_s/t_{sound}\simlt 1$ (Fig.~\ref{fig-size-rad}). { Thus, inside the shell, the dust connects} to the postshock gas flow and are { then} destroyed. Only negligible fraction of them penetrate deeper into the hot bubble and ejecta. This { can be} seen from dust density radial profiles in Fig. \ref{fig-2d-slices}, where the amount of { surviving interstellar} dust beneath the shell at $t\simgt 50$ kyr falls below 7\%. 

During the evolution the injected and { interstellar} dust remain practically isolated: the former being locked in the ejecta and the hot bubble, the latter is swept up into the shell. Only a minor fraction of the { interstellar} dust can { reach} deep in the bubble interior under conditions of weak collisional coupling $\tau_s/t_{sound}\gg 1$. Mild signs of intermixing can { be seen} only in a thin layer deeply behind the shock front, as can be seen { in the} radial distributions of dust density in Fig. \ref{fig-2d-slices}.

\begin{figure}
\includegraphics[width=8.5cm]{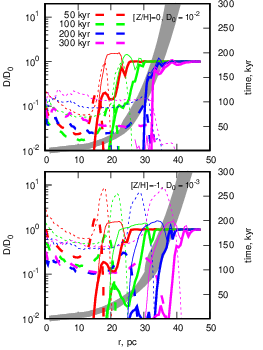}
\caption{
The radial distribution of the normalized dust density (i.e. the dust-to-gas ratio $D$ normalized to the initial background value $D_0 = 10^{-2+{\rm [Z/H]}}$) at times 50, 100, 200 and 300~kyr (color lines). Solid lines show the { interstellar} dust: thin lines represent the case without dust destruction, thick lines illustrate the case when destruction is accounted. Dashed lines depict the injected dust: thin lines show dust without destruction, whereas the thick ones are for dust with destruction being accounted. Shaded area represents the evolution of the bubble shell as defined in Fig. \ref{fig-size-rad} {\it upper} panel shows the model with ambient gas of metallicity ${\rm [Z/H]}=0$ and $\zeta_d=10^{-2}$, {\it lower} panel is for ${\rm [Z/H]}=-1$ and correspondingly $\zeta_d=10^{-3}$, the injected dust mass is $M_d=1~\msun$ in both cases. 
}
\label{fig-md-rad}
\end{figure}

The tendency of the injected and the { interstellar} dust to stay in the remnant rather unmixed can also be seen in Fig. \ref{fig-md-rad}, where the radial profiles of their abundances are given at different times. Both the { interstellar} and the injected dust-to-gas ratios $D$ are normalized to the initial value $D_0 = 10^{-2+{\rm [Z/H]}}$ of the {\it interstellar} dust. As the dust-to-gas ratio of injected dust is set $\zeta_{ej}=1/30$ (see in Sec. \ref{sec:model}), its {\it normalized} { dust-to-gas} ratio in the lower panel of Fig. \ref{fig-md-rad} for ${\rm [Z/H]}=-1$ is higher than in the upper panel for ${\rm [Z/H]}=0$. At $t=50$ kyr, a considerable fraction of the injected dust in the thermalized { bubble} is already destroyed, $\sim 0.8$. { During the following expansion over the next 150 kyr, dust destruction becomes less efficient:} at $t>100$ kyr around 8--9 \% of the injected dust particles survive, at $t\simgt 300$ kyr the injected dust practically merges to the shell and the { surviving} dust fraction falls below $\sim 5$\%; no injected dust particles penetrate beyond the forward shock. The { interstellar} dust { is efficiently destroyed behind the shock only} at early stages $t\simlt 100$ kyr, while at later stages the post-shock gas cools radiatively and temperature falls below the sputtering threshold, as discussed above, Sec. \ref{timevar}. { In both high and low metallicity gas:} ${\rm [Z/H]}=0$ and ${\rm [Z/H]}=-1$, intermixing between the injected and the { interstellar} particles is seen in the rather narrow region { inside} the bubble.  While the injected particles tend to occupy practically the entire bubble, the { interstellar dust} penetrate inwards until being { mixed} into expanding remnant gas, { since} $|v_p-v_{gas}|\ll\sigma_t$. As $\tau_{\rm sp}/\tau_s\simlt 0.3$ only a negligible amount of { interstellar} dust can penetrate deep into the hot bubble interior, resulting in a sharp drop of their abundance seen in Fig. \ref{fig-md-rad}. 

\section{Discussion and conclusion} \label{sec:discus}

{ The dust properties in the ISM can be spatially inhomogeneous due to contributions from the two dust populations: the population injected by SNe and the shock processed pre-existing one. Both depend on the local gas density and the local SN rate. For the ambient gas density $n\geq 0.1$ cm$^{-3}$ only $\simlt 13$\% of injected dust particles with radii $a_0\sim 0.1~\mu$m do survive; at higher $n$ this fraction decreases as $n^{-1.2}$. Larger particles are less sensitive to the ambient density: the surviving fraction of dust grains with $a_0\sim 0.3~\mu$m is more than $30$\% and weakly depends on ambient density $\propto n^{-0.15}$.} 

{ Destruction of the pre-existing dust particles in ambient gas is less severe. Estimates of the dust budget for a SN remnant at its radiative stage, $t\simgt 30n^{-1/3}$ kyr, show that { up to} $\sim 80$\% of the swept up { interstellar} dust survives. Therefore, after $t\geq 30n^{-1/3}$ kyr, the remnant shell contains the mass of the surviving swept up dust up to $\geq 0.8M_{d,e}^{swept}\sim 32n^{2/5}~\msun$. On the other hand, the mass of the { surviving} injected dust in the remnant is only $\leq 0.13 M_{d,i}$, which results in a ratio of injected-to-interstellar dust in the remnant at its later stages of $M_{d,i}/M_{d,e}\geq 0.004$. This impurity is too low to be detected in the swept up shell. Possible observational signatures of differences between optical properties of the pre-existing and the shock processed dust can be recognized in the interface region separating the dense SN shell and its rarefied bubble. Spatial variations of the exinction law described by \citet{Fitzpatrick1990} may be connected with contributions from such interfaces. Quite recently signatures of the two populations of dust with rather distinct spectral features are recognized in the {\it Planck} emission bands at high Galactic latitudes \citep[][ their Figs. 3, 4 \& 7]{shiu2023evidence}.}

Dust supply from SN explosions in galaxies at $z<5$ plays apparently a minor role, particularly when the small dust particles with $a<0.1~\mu$m are concerned. The survival percentage of $a\simgt 0.1~\mu$m dust { in} high density environment $n>1$ cm$^{-3}$ is less than 10\%. This estimate is qualitatively consistent with early consideration by \citet{Nozawa2006,Nozawa2007}. { The} percentage of particles of small sizes is much lower than 1\% in a $n\simgt 1$ cm$^{-3}$ environment. Our results show that the survival fraction of dust particles of $a\sim 0.1~\mu$m is $f_{d,s}\sim 10^{-2}\bar M_{d,i}n^{-\beta}$, $n$ is the ambient density in cm$^{-3}$, $\beta>1$, $\bar M_{d,i}$ is the dust mass injected by a single SN. This results in an estimate of dust supply rate due to SN 
\be 
\dot M_{\rm d}^+\sim 10^{-4}\bar M_{d,i}n^{-\beta}~{\rm SFR}~\msun~{\rm yr}^{-1},
\ee
{ where} $\bar M_{d,i}$ is in $1~\msun$, ${\rm SFR}$ is the SF rate in $1~\msun~{\rm yr}^{-1}$, for the { specific supernova} rate $\bar\nu_{sn}\sim 10^{-2}~\msun$. For comparison, the { typical SN dust} destruction rate in the Milky Way is \citep[see, e.g., in ][]{Draine2009} 
\be 
\dot M_{\rm d}^{-}\sim 0.5~{\rm SFR_5}~\msun~{\rm yr}^{-1}.
\ee 
{ This estimate seems to be applied particularly for dust production at the ``cosmic noon'' epoch, where the SNe injected dust is to be efficiently destroyed in the ejecta themselves at a higher density environment.}

SN explosions as the { the primary dust-formation channel} in galaxies can be important only for $z>5$ galaxies. Moreover, in most distant galaxies at $z>10$ only large (grey) dust can survive the extensive sputtering in SN shocks, because of their on average higher density environment $\propto (1+z)^3$. Estimates of the global dust budget in the ISM of high-redshift galaxies have to take into account highly sensitive sputtering yields on the shock velocity and ambient gas density, particularly when small dust particles are { considered}. At such conditions, a general view that SN explosions in high-$z$ galaxies are exceptionally selective with respect to a preferential destruction of small size dust particles. At such circumstances possible growth of dust particles {\it in situ} in interstellar medium, as recently discussed \citep[][]{Hirashita2011,Mattsson2020,hirashita2023submillimetre}.


\section*{Acknowledgements}

We are thankfull to Biman B. Nath for valuable comments. We would like to express gratitude to the anonymous referee for criticism and usefull suggestions that improved the paper. Numerical simulations of the bubble dynamics were supported by the Russian Science Foundation (project no. 23-22-00266). YS acknowledges the hospitality of the Raman Research Institute.

\section*{Data Availability}

The data underlying this article are available in the article.

\bibliographystyle{mn2e}
\bibliography{p-bib}

\appendix

\section{Dynamics of dust particles}

Here we describe an implementation of the particle dynamics in our gasdynamic code \citep{vns2015,vsn2017,vsn2019}. This code has successfully passed the whole set of tests proposed in \citet{Klingenberg2007}. Several additional tests have been given in the appendix of \citet{vsn2017}.

Our description of dust dynamics basically follows to the method proposed by \citet{Youdin2007}, several parts are similar to \citet{Mignone2019} and \citet{Moseley2023}. { W}e add several points allow to discriminate sorts of particles, to follow the evolution of dust (macroscopic) mass. { W}e include destruction processes due to thermal and kinetic sputtering. { Since} we are going to study gas-dominated (by mass) flows in the ISM, we use explicit methods for solving the equations of dust dynamics.

\subsection{Properties}\label{prop}

Dust particles trace the motion of grains in a gas. They are suffered by drag force from a gas, and due to it they transfer momentum and heat to a gas. Also they may be exposed to other external forces like gravity, radiation and so on. 

Each dust particle is described by several features allow one to identify its evolution and physical properties, namely, colour, time of injection, sort, size and mass. Two former can be used to identify a source of { each} particle. Dust particles can consist of different material -- a sort of particle, e.g, carbon or silicate. They can be of various sizes, which are distributed according to  { some} spectrum or have a single size. In the former case the dust is initially polydisperse and one can choose several sizes distributed by { some} spectrum. In the latter it is monodisperse, but due to destruction processes it { becomes} polydisperse. These features are microscopic. 

To follow the transport of dust mass (not of an individual grain) in a gaseous flow we introduce a macroscopic mass of a 'superparticle' or macroparticle. In this approach a particle is a conglomerate of microscopic grains. This value of mass (or supermass) is used to find the dust-to-gas mass ratio and so on. If it is necessary, any feature of dust particle can be added.

\subsection{Dynamics}\label{dyn_def}

The dust component is modelled as an ensemble of macroparticles governed by the system of ODEs:
\ba
 {d\pmb{x}_p \over dt} = \pmb{v}_p \\
 {d\pmb{v}_p \over dt} = \pmb{a}_p - {\pmb{v}_p - \pmb{v}_g \over \tau_s}
\label{dust-evol}
\ea
where $\pmb{x}_p$ and $\pmb{v}_p$ are the dust particle position and velocity vectors, $\pmb{a}_p$ is the acceleration vector of external forces (excepting gas-grain drag force), the stopping time is written for the Epstein drag with the supersonic Baines correction \citep{Epstein1924,Baines1965,Draine1979b}:
\be
 \tau_s = {m_p \over \pi a^2 \rho_{gas} \xi |v_p - v_{gas}|},
\ee
where $m_p$, $a_p$ and $v_p$ are the mass, size and velocity of a dust particle, respectively, $\rho_{gas}$ and $v_{gas}$ are the density and velocity of gas, and the correction is 
\be
 \xi = \left[1 + {128 k_B T_{gas} \over 9 \pi m_H (v_p - v_{gas})^2}\right]^{1/2}
\ee
where $k_B$ is the Boltzmann constant, $T_{gas}$ is the temperature of gas, $m_H$ is the mass of the proton.

Due to the stopping time may be short enough compared to the gasdynamic timestep the system becomes stiff and then one should use implicit or semi-implicit schemes for solving it \citep[see e.g.,][]{Vaidya2018,Moseley2023}. This controls by the relation between the  gasdynamic time (including the cooling time if necessary) and the stopping time, in the Epstein regime it is easily written as
\be
 \tau_s \simeq {\rho_m a \over \rho_g \sigma_T}
\ee
where $\rho_m$ is the density of a solid particle's material, $a$ is the size of a particle, $\rho_g$ is the density of the gas, $\sigma_T$ is the mean thermal velocity. So that the transition to the stiff ODEs depends on the physical conditions and aims of simulation. In the diffuse interstellar medium with number density of 1~cm$^{-3}$ and temperature of $10^4$K the stopping time of small particles with $a\simlt 10^{-6}$cm is around $10^5$yrs. In a hot gas of SN interiors it drops to quite short value of $\sim 10^2-10^3$yrs for $n_g \sim 10^{-1}-10^{-2}$cm$^{-3}$ and $T\sim 10^8$K at early free expansion phase. This time scale is around the Courant time for the cell of 1~pc with the same physical conditions. Then, one can assume that such small particles are coupled with gas when its stopping time is shorter than the gasdynamic timestep. More exactly, small particles with short enough stopping time follow the gas as Lagrangian trace particles. In this approach the system of ODEs for the particle dynamics can be solved using the explicit 'predictor-corrector' stencil adopted in the code. 

For particles we have implemented periodic, outflow and reflective boundary conditions. They are similar to those for gas.

\subsection{Back-reaction to the gas}

The drag force of a grain is proportional to the relative velocity between gas and particles \citep{Weiden1977}:
\be
 \pmb{f}_d = -\rho_d {\pmb{v}_g - \pmb{v}_d \over \tau_s}
 \label{dragforce-m}
\ee
where $\pmb{v}_g$, $\pmb{v}_d$ are the velocities of gas and dust particle, $\tau_s$ is the stopping (friction) time, for which we have implemented several regimes: the constant stopping time, the Epstein drag, and the Epstein drag with the supersonic Baines correction \citep{Epstein1924,Baines1965,Draine1979b}. Usually we use the latter.

In the total energy equation we need to take into account the work done by the drag force and the frictional heating due to dust:
\be
 \pmb{v}_d \pmb{f}_d = \pmb{v}_g \pmb{f}_d + \rho_d {(\pmb{v}_d - \pmb{v}_g)^2 \over \tau_s }
 \label{dragforce-e}
\ee

The momentum and energy equations are added by source terms responsible for the back-reaction to the gas, see eqns~\ref{dragforce-m}-\ref{dragforce-e}:
\be
 {\partial (\rho_{gas}\pmb{v}_{gas}) \over \partial t} + ... = ... + \pmb{f}_d,
\ee
and
\be
 {\partial E_{gas} \over \partial t} + ... = ... + \pmb{v}_d \pmb{f}_d. 
\ee

\subsection{Destruction}\label{app_dest}

{ While moving through a gas dust particles can be destroyed} by both thermal and kinetic sputtering \citep{Draine1979b}. We use the thermal sputtering rate from \cite{Draine-book} and the kinetic one from \cite{Nozawa2006}. For the latter we approximate the  rate presented by \cite{Nozawa2006} in their Fig. 2b, averaged over the target species: 
\be
\dot a = 4\times 10^{-4} n_{gas}  {\rm exp}\left(-{4\times 10^2 \over |v_p -v_{gas}|}\right) |v_p -v_{gas}|^{-0.7}$~$\mu$m~yr$^{-1},
\ee
where the velocities are in km s$^{-1}$.

For rough estimates of nonthermal destructions one can utilize the following equation for supersonic relative dust motion \citep{Draine1995} 
\be 
4\rho_m{da\over dt}=-n m_t w Y(w), 
\label{ntsput}
\ee 
{ where} $m_t$ is the target mass ($m_t\simlt 20-30m_p$), $w\equiv v_p-v_g$. Dividing Eq. (\ref{ntsput}) over Eq. (\ref{dust-evol})  { results in}   
\be 
{d\ln a\over d\ln w}={m_t Y(w)\over 3 m_{\rm H}}, 
\ee 
{ or} approximately 
\be 
a\geq a_0\left({w\over w_0}\right)^q
\ee  
with $q=m_t \langle Y(w)\rangle/3 m_{\rm H}$, $\langle Y(w)\rangle\simeq 0.01$ being the mean value of the { sputtering} yield $Y(w)$ averaged over the target species in \cite{Nozawa2006} in the velocity range of interest $w=100-500$ km s$^{-1}$. This results in $a\geq 0.88a_0$, or for mass depletion $m(a)\geq 0.7m(a_0)$.

For injected dust particles the kinetic sputtering seems to be even weaker because they seem to be involved into the ejecta flow at initial ``ejecta-dominated'' stages, with { low relative velocities of dust and gas and a short} $\tau_s\simlt 1$ kyr in the ejecta, as mentioned in Sec. \ref{sec:model}.

\subsection{Grid and particle's quantities}
\label{sec-deposit}

For depositing a particle quantity $q_p$ {\bf(e.g., velocity)} to the grid and interpolating fluid quantity $Q_{ijk}$ at the particle location we use
\be
 Q_{ijk} = \sum_{p=1}^{N_p} {W(\pmb{x}_i-\pmb{x}_p)q_p}, \hspace{0.3cm}  q_p = \sum_{ijk} {W(\pmb{x}_{ijk}-\pmb{x}_p)Q_{ijk}}, 
\ee
where the kernel $W(\pmb{x}_i-\pmb{x}_p) = W(x_i-x_p)W(y_i-y_p)W(z_i-z_p)$ is product of three one-dimensional weight functions { \citep[see, e.g.][]{Birdsall1991}}, and only neighbour cells give a nonzero contribution to the particle in the second sum. We have implemented traditional shape functions such as the ``Nearest Neighbour Point'' (NGP), ``Cloud-In-Cell'' (CIC), and ``Triangular Shape Cloud'' (TSC), we have adopted that each particle can give a nonzero contribution only to the computational zone hosting the particle, its left and right neighbours. 

\subsection{Memory allocation}

In our realization particles are held in a computer memory using a singly linked list consisting of sequentially linked structures. Each node contains the information about the current particle and the pointer to the next node in the sequence. In this list nodes can be inserted or removed. A particle is described by several individual features such as identical number, three coordinates, three velocities, time moment of injection and so on. 

\section{Tests}\label{sec:gentest}

\subsection{Particle-gas deceleration test}

We study the dynamical interaction of an ensemble of uniformly distributed dust particles with a homogeneous gas flow. Initially particles are placed { at rest} into a flow with a constant velocity $v_{g,0}$. Assuming fixed stopping time for particles we get an analytic solution for the evolution of particle velocity: $v_p(t) = v_{g,0} \left[1 - {\rm exp}(-t/\tau_s)\right]$. 

\begin{figure}
\includegraphics[width=8.5cm]{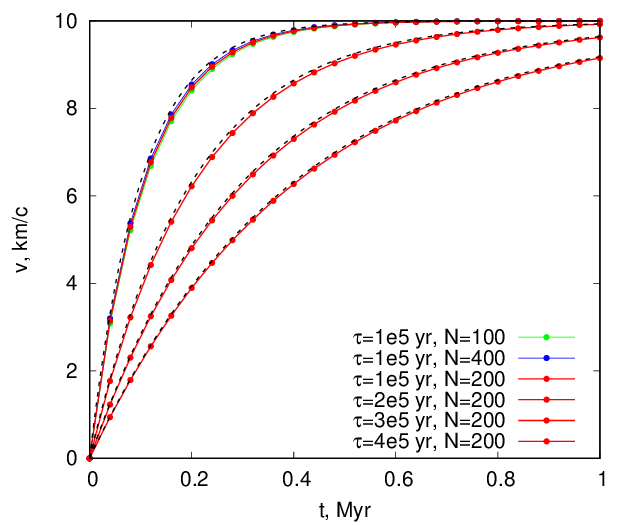}
\caption{
The velocity of dust particles for stopping times $\tau_s = 10^5$, $2\times 10^5$, $3\times 10^5$ and $4\times 10^5$yr (red solid lines from top to bottom, respectively) in a homogeneous gas flow for grid with number of cells along each spatial direction $N=200$. The green and blue lines show the velocity for $\tau_s = 10^5$yr and number of cells $N=200$ and 400. The black dash lines represent the analytic solution for stopping times $\tau_s = 10^5$, $2\times 10^5$, $3\times 10^5$ and $4\times 10^5$yr (from top to bottom). 
}
\label{fig-d-test}
\end{figure}

In our test runs of the coupled particle-gas dynamics we set the gas flow velocity equal to $v_{g,0}=10$~km/c, and investigate the velocity of particles for different stopping times. Figure~\ref{fig-d-test} presents the particle velocity for $\tau_s = (1-4)\times 10^5$yr for grid with number of cells along each spatial direction $N=100$, 200 and 400 (colour lines). One can see a good coincidence between our numerical results and the above-mentioned analytic solution (depicted by black dashed lines). Only for the coarse grid ($N=100$) and the shortest stopping time one can { find} small deviations from the analytic curve. 

\subsection{SN evolution: a convergence}
\label{sec-sntest}

\begin{figure}
\includegraphics[width=8.5cm]{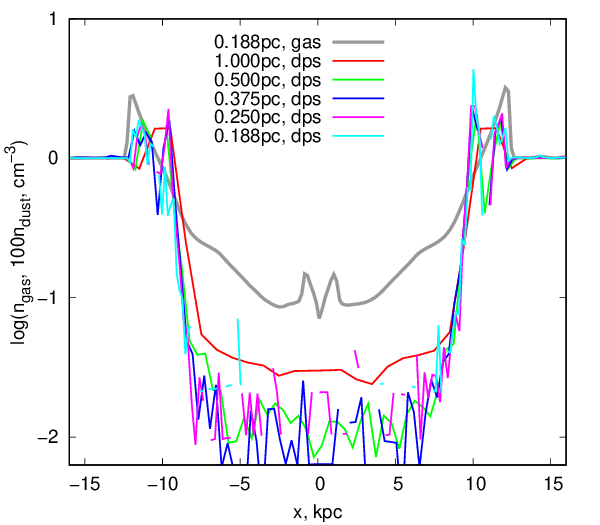}
\caption{
The one-dimensional profile of the dust density depositing to the grid for different spatial resolution of the grid: the colour lines correspond to the cell size from 1 to 0.1875~pc. The slice is for a SN bubble with age of 20~kyr. The number of injected particles in all models is equal to 8 millions. The thick grey line depicts the gas density profile along the same line of sight and for the highest spatial resolution (0.1875~pc). The dust density is multiplied by a factor of 100.
}
\label{fig-n-1d-res}
\end{figure}

We follow the evolution of a single SN remnant with different spatial resolution for the grid and fixed number of dust particles injected by SN. The number of dust particles in the ambient gas (or { pre-existing} dust particles) is proportional to the grid resolution, because we set one such particle per grid cell. Figure~\ref{fig-n-1d-res} presents the one-dimensional (along line of sight crossed the center of SN bubble) profile of the dust density depositing to the grid for different spatial resolution of the grid: from 1 to 0.1875~pc. The slice is for a SN bubble with age of 20~kyr. The number of injected particles remains the same and equal to 8 millions. For comparison the gas density profile along the same line-of-sight and for the highest spatial resolution (0.1875~pc) is added. The dust density is multiplied by a factor of 100 for better presentation (the dust-to-gas density { ratio} in the ambient gas is initially set to 0.01). One can see that the dust distributions { for the cell size lower than 0.5 pc are close}. { Therefore,} we conclude that the grid resolution of 0.5~pc is sufficient to follow the dust dynamics during the SN bubble evolution. { The gaps in profiles seen for high resolution are due to a limited number of particles in a cell.} 

\begin{figure}
\includegraphics[width=8.5cm]{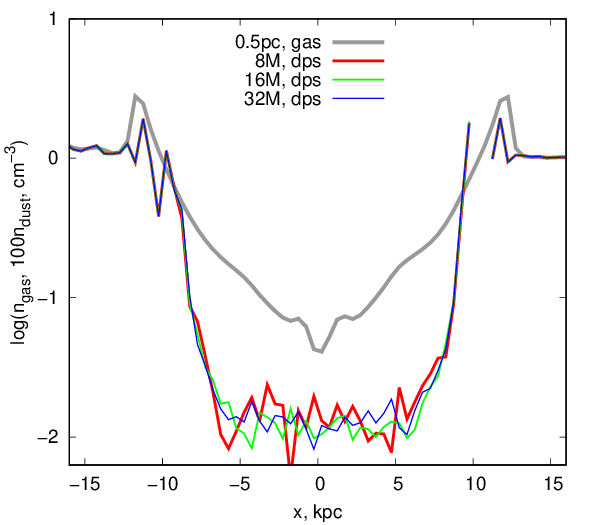}
\caption{
The one-dimensional slice of dust density for the runs with different number of dust particles: the colour lines show the profiles for 8, 16 and 32 millions of particles. The grid resolution is the same and equal to 0.5~pc. The thick grey line depicts the gas density profile for this resolution. The age of a SN bubble is 20~kyr. The dust density is multiplied by a factor of 100.
}
\label{fig-n-1d-ndps}
\end{figure}

Figure~\ref{fig-n-1d-ndps} shows the one-dimensional distribution of dust density for the runs with different number of dust particles: 8, 16 and 32 millions, but the { fixed} grid resolution equal to 0.5~pc. The slice is for a SN bubble with age of 20~kyr. { It is seen that the dust density profiles} { for the runs with different number of particles are very close}. { The number of dust particles decreases within the bubble faster than the gas density, because of decoupling in  the innermost regions of the bubble.}  { Therefore, for our runs we adopt the spatial resolution equal to 0.5~pc, and the number of injected dust particles  equal to 8 millions as fiducial. These values are sufficient to follow the dust dynamics inside a SN bubble adequately.}

\section{Lagrangian particles}

We have implemented in our code Lagrangian or tracing particles, which follow the gaseous flow and mark regions of a gas filled by injected 'fluid' without diffusion.

\subsection{Properties}

In addition to several general features of particles needed to the identification and evolution each Lagrangian particle is described by its colour and time of its birth or injection. For instance, a colour can be used to identify a source of { each} particle.

\subsection{Dynamics}

The dynamics of Lagrangian particles is modelled as an ensemble of particles governed by the system of ordinary differential
equations (ODEs):
\be
 {d\pmb{x}_p \over dt} = \pmb{v}_g
\label{trace-evol}
\ee
where $\pmb{v}_g$ is the gas velocity vector in the cell where a particle is located.

We solve this system coupled with the gasdynamic equations using the predictor-corrector scheme realized in the gasdynamic code.

For depositing a particle quantity $q_p$ to the grid and interpolating fluid quantity $Q_{ijk}$ at the Lagrangian particle location we use the same description as for dust particles.

\end{document}